\documentclass{elsart}
\usepackage{natbib,graphicx,amssymb}

\def\re{\mbox{$r_e$}}

\def\muem{\mbox{$\langle \mu \rangle_e$}}
\def\ie{\mbox{$I_e$}}
\def\iem{\mbox{$\langle I \rangle_e$}}

\def\mue{\mbox{$\langle \mu \rangle_{\rm e}$}}

\def\muere{\mbox{$\langle\mu\rangle_{\rm e}$-$\log(r_{\rm e})$}}
\def\H0{\mbox{$H_0$}}
\def\q0{\mbox{$q_0$}}

\def\r1q{\mbox{$r^{1/4}$}}
\def\rn{\mbox{$r^{1/n}$}}

\def\gradip{\hbox{\rlap{\hbox{.}}\raise 5.truept \hbox{{\small $\circ$}}}}
\def\magcir{\ \raise-2.truept\hbox{\rlap{\hbox{$\sim$}}\raise5.truept
    \hbox{$>$}\ }}
\def\etal{et al.\/}

\def\ie{{i.e.\/}}
\def\eg{{e.g.\/}}

\def\mue{\mbox{$\langle \mu \rangle_{\rm e}$}}

\newenvironment{lefteqnarray}{\arraycolsep=0pt\begin{eqnarray}}
{\end{eqnarray}\protect\aftergroup\ignorespaces}
\newenvironment{lefteqnarray*}{\arraycolsep=0pt\begin{eqnarray*}}
{\end{eqnarray*}\protect\aftergroup\ignorespaces}

\begin{document}
\begin{frontmatter}
\title{Toward Understanding the origin of the Fundamental Plane for Early-Type Galaxies.}
\author{M. D'Onofrio}
\ead{donofrio@pd.astro.it}
\author{T. Valentinuzzi}
\ead{valentinuzzi@pd.astro.it}
\author{L. Secco}
\ead{secco@pd.astro.it}
\author{R. Caimmi}
\ead{caimmi@pd.astro.it}
\author{D. Bindoni}
\ead{bindoni@pd.astro.it}
\address{Astronomy Department, Padova University, Vicolo dell'Osservatorio 2, 35122 Padova, Italy}

\begin{abstract}
We present a panoramic review of several observational and
theoretical aspects of the modern astrophysical research about the
origin of the
Fundamental Plane (FP) relation for Early-Type Galaxies (ETGs).
The discussion is focused on the
problem of the {\it tilt} and the {\it tightness} of the FP, and on the attempts
to derive the luminosity evolution of ETGs with redshift.
Finally, a number of observed features in the FP are
interpreted from the standpoint of a new theoretical approach based
on the two-component tensor virial
theorem.

\end{abstract}

\end{frontmatter}

\section{INTRODUCTION}
The long-debated question of how many physical independent parameters
can be used to describe the overall observational manifold of
galaxies, arose approximately 30 years ago, in particular with a
pioneering paper by \citet{aa23-259}. Performing a principal component
analysis (PCA) on a sample of spirals with known rotation curves,
morphological types, absolute luminosities, colors and [HI] masses, a
surprising result arose, that only 2 independent parameters dominate
most of the variance of the related manifold. \citet{brosche88} proved
the possibility to scale all the main integral galaxian properties
with two parameters, \eg\ mass and angular momentum. Since then, the
multivariate analysis on galaxian parameters has been carried out by
many authors \citep[e.g.,][]{Lentes,mnras206-453,apj278-61,apj280-7},
substantially confirming the earlier conclusion.  An application of
PCA to early-type galaxies (ETGs) yielded a similar result
\citep[][]{BRLen}, providing additional support to two previously
discovered correlations, involving residuals from the Faber-Jackson
(FJ) relation, $L_T \propto \sigma_o^{J}$ \citep[e.g.,][]{apj204-668}.
More precisely, the correlations are between (i) residuals in total
luminosity, $L_T$, and metallicity index, $Mg_2$
\citep[][]{aj100-1416}, and (ii) residuals in central projected
velocity dispersion, $\sigma_o$, and mean effective surface
brightness, \muem\ \citep[][]{apj230-697}.

At that time, the idea that ETGs belong to a family of stellar
systems controlled by a single parameter, where their physical
properties scale according to luminosity (mass), was supported by several observational evidences: the large degree of
homogeneity of galaxy light profiles (well represented by the
\r1q\ de Vaucouleurs law), the uniform color-magnitude diagrams, the
existence of the FJ relation and of the Kormendy relation \citep[KR,][]
{Korm77}, $\muem=3\log(r_e)+const$, between the effective radius,
\re\, and \muem.

The above simple scenario started to change when \citet{apj313-59} and
\citet{apj313-42}, taking advantage of a large sample of ETGs with
available photometric and kinematical data, derived simultaneously the
observational evidence that three physical parameters, $\sigma_o$,
\muem, and \re , are mutually correlated and unified in a
2-dimensional manifold, since then called {\it Fundamental Plane}
(FP).
In this framework, both FJ and KR were interpreted as
projections of the FP along the coordinate axes, which provides a natural
explanation to the second, mysterious, ``hidden parameter''.

Current astrophysical observations and theoretical
speculations have widely extended the spectrum of the structural and
dynamical parameters that can be measured or calculated within
galaxies, including both
the visible stellar (baryonic) component, B, and the invisible dark
matter (DM) component, D.
Despite this
large zoo of parameters, it is very surprising that the main dimensionality
of this manifold remains 2.
In other words, the cloud of points in
the parameter space whose axes are size (mass, luminosity, or radius),
density (or surface brightness) and temperature (\ie\ kinetic energy
per unit mass), does not populate uniformly the three dimensional
space, but is distributed approximately along a plane where the
scatter maintains small.

The current paper aims to briefly review the most relevant attempts to
explain the origin of the FP for ETGs. The mere existence of the FP
does indeed indicate that structural properties in ETGs span a narrow
range, suggesting some self-regulating mechanism must be at work
during formation and evolution. The most relevant observational
features of the FP are the tilt with respect to what is expected from
one-component scalar virial theorem and homology (see section \ref{sec1}), and
a small thickness called tightness (see section \ref{sec2}).
According to observations, the scatter around the FP is very low, and
the position of a galaxy above or below the plane is independent of
galaxy flattening, isophotal twisting, velocity anisotropy, and details
of radial light distribution. The small thickness corresponds to
about 12\% uncertainty in \re , implying the FP is a good distance
indicator \citep[e.g.,][]{mnras330-443}.

The paper is organized as follows. In section \ref{sec1} we review the most relevant works that in the last years
provided new data for the FP at different wavelengths, redshifts and
environments. Special effort is devoted to both the KR and the $\kappa$-space. The latter makes a very interesting and debated representation of
the FP, built up by \citet{apj399-462} using a different set of
orthogonal variables. In section \ref{sec2} we summarize the
most important theoretical attempts devoted to
explain the origin of both the tilt and the tightness of the FP.
We will address, in particular, the role of
anisotropy and rotation in the stellar velocity distribution, the weak
deviation of stellar systems from homology, the role of
Initial Mass Function (IMF), the role of DM, and the most relevant
aspects of a new
theoretical approach which uses the two-component tensor virial theorem
for interpreting some of the observed features of the FP.
The conclusions are drawn in section \ref{sec4}.

\section{The observed properties of the FP}\label{sec1}
The most diffuse representation of the FP writes:
\begin{lefteqnarray}
\label{eq:pianofond}
&&\log (r_e) =  A\cdot \log(\sigma_o) + B\cdot \log(\langle I \rangle_e) + C~~;
\end{lefteqnarray}
where $A$, $B$ and $C$, are constant coefficients (for each wavelength
filter bandpass) derived by means of a multiple regression fit of the
effective radius, \re\ (encircling half the total
luminosity), the central projected velocity dispersion, $\sigma_o$, and
the average effective surface brightness in flux units, $\langle I
\rangle_e$\footnote{The exact definition of
  these parameters is not always univocal in the astronomical
literature but is not discussed here. The reader may look at
many of the papers cited in the references.}.
In table \ref{coefAB} we have listed the values of $A$ and $B$ found in several works appeared in literature.
\begin{table}
\begin{center}
\begin{tabular}{|l|c|c|c|} \hline
      & $\bf{A}\pm \bf{\Delta A}$ & $\bf{B}\pm \bf{\Delta B}$ & \bf{Band} \\
\hline
\citet{apj313-42} & $1.33\pm0.05$ & $-0.83\pm0.03$ & B \\
\hline
\citet{apj313-59} & $1.39\pm0.14$ & $-0.90\pm0.09$ & B \\
\hline
\citet{lucey} & $1.27\pm0.07$ & $-0.78\pm0.09$ & V \\
\hline
\citet{guzman} & $1.14\pm0.07$ & $-0.79\pm0.07$ & V \\
\hline
\citet{apj453-17} &$1.44\pm0.04$ & $-0.79\pm0.03$ & K'\\
\hline
\citet{mnras280-167} & $1.24\pm0.07$ & $-0.82\pm0.02$ & r \\
\hline
\citet{mnras291-488} & $1.38\pm0.04$ & $-0.82\pm0.03$ & R \\
\hline
\citet{scodeggioetal} & $1.25\pm0.02$ & $-0.80\pm0.02$ & I \\
\hline
\citet{scodeggio97} & $1.55\pm0.05$ & $-0.80\pm0.02$ & I \\
\hline
\citet{aspc116-154} & $1.66\pm0.09$ & $-0.75\pm0.06$ & K'\\
\hline
\citet{aj116-1606} & $1.53\pm0.08$ & $-0.80\pm0.02$ & K \\
\hline
\citet{apj531-184} & $1.31\pm0.13$ & $-0.86\pm0.10$ & V \\
\hline
\citet{gibbons2001} & $1.37\pm0.04$ & $-0.82\pm0.01$ & R \\
\hline
\citet{aj125-1866} & $1.49\pm0.05$ & $-0.75\pm0.01$ & r \\
\hline
\end{tabular}
\vspace{0.3cm} \caption[]{Several observational data on the
coefficients, $A$ and $B$,
  in different passbands taken from many authors.}
\end{center}
\label{coefAB}
\end{table}

\begin{figure}[!h]
\begin{center}
\includegraphics[bb=7cm 9cm 15cm 30cm,scale=0.7]{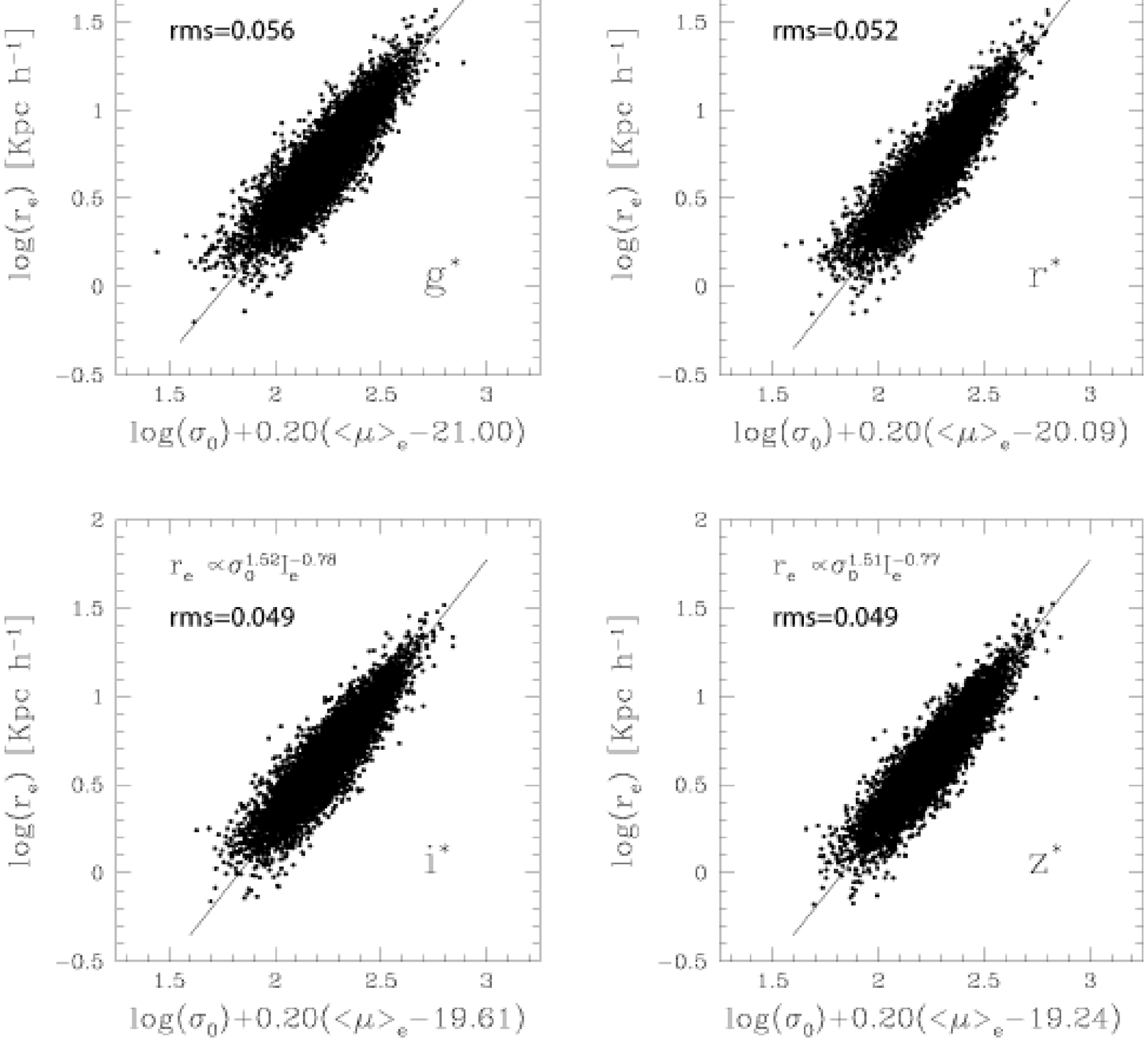}
\caption{The FP in 4 filter bandpasses for a sample of $\sim9000$
ETGs
  of the SLOAN digital sky survey \citep[for further details see,][]{aj125-1866}.}
\label{fundplane}
\end{center}
\end{figure}

The remarkable fact is a substantial variation of $A$ in the various
filter bandpasses, as opposed to an almost constant value of $B$. The
wavelength dependence of $A$ is also clearly visible in
Fig.\,\ref{fundplane}, which shows the distribution of about
$\sim9000$ ETGs extracted from the first release of the SLOAN digital
sky survey \citep[see][]{aj125-1866}. The values of $A$ and $B$ do not
agree with what is expected from one-component scalar virial theorem
and homology, which makes the well known problem of the {\it tilt} of
the FP.

For a virialized one-component stellar system, the scalar virial theorem
reads:
\begin{lefteqnarray}
\frac{GM}{R_g}=\langle\sigma^2\rangle~~;
\end{lefteqnarray}
where $M$ is the mass, $R_g$ the gravitational radius \footnote{This
definition appears ambiguous, as it is often used for the
Schwartzschild radius $R_g=GM/c^2$. A better definition appears to be
the following. Given a bound matter distribution with total mass, $M$,
and gravitational potential energy, $\Omega$, the isothermal radius,
$R_i$, is the radius of a truncated isothermal sphere with equal mass
and gravitational potential energy, $R_i=-GM^2/\Omega$.},
$\langle\sigma^2\rangle$ the mean square velocity dispersion of the
stellar system, and $G$ the gravitational constant. If ETGs make an
homologous family (\ie\ they have the same mass distribution and the
same kinematics) the following relationships between theoretical and
observational quantities can be assumed:
\begin{lefteqnarray}
R_g=\frac{r_e}{k_R}~~,~~\langle \sigma^2 \rangle =\frac{\sigma_o^2}{k_V}~~.
\end{lefteqnarray}
Accordingly, $M=c_2\sigma_o^2r_e$, where the parameter, $c_2=(Gk_Rk_V)^
{-1}$, maintains the same value for all the members of
the whole family.  Furthermore, if the total luminosity reads
$L=c_1\langle I \rangle_e r_e^2$, where the parameter, $c_1$,
maintains the same value for all the members of the whole family,
the $M/L$ ratio can be expressed as:
\begin{lefteqnarray}
\frac{M}{L}=\frac{\sigma_0^2r_e c_2}
{\iem r_e^2c_1}~~;
\end{lefteqnarray}
which is equivalent to:
\begin{lefteqnarray}
\label{eq:tiltvirial}
r_e=c_2c_1^{-1} \left( \frac{M}{L} \right)^{-1}\sigma_0^2\iem^{-1}~~.
\end{lefteqnarray}
With regard to the assumption of both homology and constant $M/L$,
Eqs.\,(\ref{eq:pianofond}) and (\ref{eq:tiltvirial}) imply $A=2$ and
$B=-1$.

On the other hand, significantly smaller values of $A$ and $B$
are deduced from Tab.\ref{coefAB}, in contradiction with previous
predictions.

\subsection{The wavelength dependence of the FP}\label{sec1a}

Early works on the FP argued that the deviation of the slope
from model predictions is due to a weak systematic variation of
the $M/L$ ratio of galaxies with their luminosity (or mass).  It is a
stellar population effect, either in age, metallicity or IMF that has
a distinctive signature in the variation of the FP slope with
wavelength.  According to \citet{aj116-1606} the slope, $A$, of the FP
relation in Eq.(\ref{eq:pianofond}), increases systematically with
wavelength from U band to K band.

An inspection of Tab.\ref{coefAB} shows that effective radii in the
Near-IR are always smaller than in the optical band, the difference in
the FP slope, $\Delta A_{({\rm U}-{\rm K})}$, being approximately
$0.5$.  This result is quite robust (even if the dependence of the FP
slope on wavelength is quite small), since it comes from a direct
comparison of same galaxies observed in different color bands. The FP
slope is obtained either by means of a direct comparison of the fits,
either by comparing the more stable quantity $log(r_e)-B'\cdot \muem$,
where $B=-2.5 B'$ due to $\muem = -2.5\log\iem$.

The above variation is commonly interpreted in terms of systematic
differences in the stellar content along the sequence of ETGs.  Color
gradients resulting from metallicity gradients may be at the origin of
this behavior, but unfortunately current models of star formation
(SF) and evolution cannot help enough to this respect.

A similar dependence of the FP slope parameter, $A$, on wavelength has
been confirmed by \citet{mnras301-1001}. Both groups observe that the
$B'$ parameter is very stable with wavelength and equal to about
0.32. \citet{mnras291-488} find $B'=0.326\pm0.011$ in $R_C$, while
\citet{mnras280-167} report $B'=0.328\pm0.008$ in Gunn-r.  The
wavelength dependence of $A$ has been also confirmed by the analysis
of the data of the SLOAN digital sky survey \citep[][]{aj125-1866}.

The above results show that the FP slope is inconsistent with
predictions at all wavelengths.  The deviation of the Near-IR FP slope
from these expectations (virial theorem + homology + constant $M/L$)
implies that at least one among the assumptions has to be dismissed.
More specifically, either the $M/L$ varies from galaxy to galaxy along
the FP sequence, or ETGs do not belong to a homologous family.

The advantage of considering the Near-IR band lies in a small
dependence on metallicity effects, and then little influence from
age-metallicity degeneracy.  The modeling proposed by
\citet{aj116-1606} predicts an age and metallicity variation along the
ETGs sequence, assigning larger age and lower metallicity to more
luminous galaxies and vice versa. The above finding seems to be in
contradiction with the theory of hierarchical merging
\citep[e.g.,][]{mnras281-487}, in which present-day galaxies result
from successive merging of small units.

The lack of correlation between the residuals of the Near-IR FP with
those of the $Mg_2-\sigma_o$ relation indicates, that the tightness of
the FP cannot be ascribed only to age or metallicity effects
\citep[][]{asp163}.

Both the age-metallicity model of \citet{Worthey95} and the dynamical model
of \citet{apj451-525}, based on the deviation from homology in the
velocity distribution of ETGs, cannot explain the variation of
FP slope with wavelength.

\subsection{The FP at high redshift}\label{sec1b}
The FP relation, together with the Tully-Fisher relation and the FJ
relation, have often been used to constrain the mass evolution of
galaxies at increasing redshift. The small scatter of the above
mentioned relations
implies a small spread in age, approximately of $20$-$30\%$, and
makes them very useful for evolutionary effect analysis.

Studies on evolutionary effects are possible thanks to the high
efficiency of the Hubble Space Telescope and the high spectral
resolution of big ground based telescopes. The above mentioned
investigations aim to discriminate between two traditional galaxy
formation and evolution scenarios, namely (i) the monolithic collapse
model \citep[e.g.,][]{apj136-748,apj179-427}, in which galaxies form
at very early epochs from the collapse of a protogalactic gas cloud,
implying little evolution of the mass function after $z\simeq3$, and
(ii) the hierarchical model in which galaxies continuously form in
mergers and the mass function changes in the redshift range, $0\leq z
\leq 2$ \citep[e.g.,][]{mnras264-201}.  It may safely be expected that
inner and denser regions of proto-galaxies underwent something like a
monolithic collapse, while outer and less dense regions were accreted
via hierarchical clustering.

By exploiting the FP, \citet{mnras281-985} derived the evolution of the
$M/L$ ratio from Coma ($z=0.023$) to the cluster CL0024
($z=0.39$). Their data suggest a decrease of $31\pm12\%$ in
$M/L$ as $z$ increases. A similar trend is expected for the evolution of
stellar populations. The luminosity of a stellar population with
fixed mass, is expected to change in time as:
\begin{lefteqnarray}
L \propto 1/(t-t_{\rm form})^k~~;
\end{lefteqnarray}
where $t_{\rm form}$ is the formation epoch. To the first order,
$k=1.3-0.3x$, where $x$ is the IMF slope, but $k$ depends also
on wavelength and metallicity. Models by \citet{apjs71-817} and
\citet{apjs95-107} produce $k=0.6$ and $k=0.95$ in V band, for
various IMF and metallicities.  As a result of $M/L$ evolution, the
zero point of the FP is expected to change, and if the variation in
$M/L$ depends on $M$, the coefficients of the FP must also change.  The
evolution of $M/L$ as a function of redshift, can be expressed as:
\begin{lefteqnarray}
\ln \frac{M}{L}(z) = \ln \frac{M}{L}(0) - k(1+q_0+z_{\rm{{\rm form}}}^{-1})z~~;
\end{lefteqnarray}
where $z_{\rm form}$ is the formation epoch of the stellar population,
and $q_0$ is the cosmological deceleration parameter. In general, the
$M/L$ ratio decreases with redshift as the luminosity increases.

If galaxies evolve passively, \ie\ without merging or accretion, then
the product, $\sigma_o^{0.49}r_e^{0.22}$, should remain constant and
only the $M/L$ ratio will evolve in time.  A great accuracy is
therefore required in determining the tilt of the FP at large $z$.
Unfortunately, the small sample of galaxies generally observed at high
redshift (typically around 10 objects) does not permit a rigorous
comparison between FP coefficients at low and high $z$. This is why
the sample does not include galaxies of low masses, at high $z$.
The difference in the mean mass of galaxy samples results
in a strong correlation between the zero point offsets and the tilt
values of the FP.

A further effect is that the luminosity function of ETGs does not
necessarily evolve in the same way as the $M/L$ ratio.  If
dissipationless merging occurs frequently, the luminosity of the final
product is affected in a stronger way with respect to the $M/L$ ratio.
For a correct interpretation of the luminosity function, it is
essential to measure $M$ and $M/L$ independently. Total masses are
notoriously difficult to measure, as scaling relations are
used to constrain the evolution of $M/L$.

Another possible bias has been pointed out by
\citet{IAUS164-269F}: the fraction of ETGs in clusters may
evolve with redshift. Accordingly the set of ETGs at
low redshift may be different from the set observed at high $z$. Some of
the present-day ellipticals may in fact be spirals at large $z$. A
measure of the morphological evolution would be required to take into
account the above mentioned effect.

The determination of the luminosity evolution, $k$, is probably the most
important difficulty. Its dependence on IMF is a serious problem, since
the IMF of ETGs is not well determined yet. The color evolution
of galaxies is also linked to both IMF and $k$. For a
steep IMF $k$ is small, but the color evolution is rapid \citep[see][]{Tinsley1980}.

Finally, another fundamental assumption has been made, namely a
cluster at high $z$ will evolve in the cluster at low $z$ used for the
comparison. In this view, a cluster observed at higher $z$ has to
resemble as much as possible the real progenitor of the comparison cluster observed at
lower $z$. The richness class may play a role in determining the
properties of the FP.

In general, the data acquired by several authors in the field of
research discussed above, seem to confirm that massive ellipticals
existed at high redshifts and were very similar to present-day objects
in nearby clusters. A passive galaxy evolution is very consistent with
observations.

The methodology under consideration has been extended up to $z=1.27$
by \citet{apj585-78}. The results are again consistent with the
passive luminosity evolution. For galaxies with masses greater than
$10^{11} {\rm M}_{\odot}$, the following trend results: $\ln (M/L_B)
\propto (-1.96\pm0.09)z$, corresponding to a passive evolution of
$-1.50\pm0.13$ mag at $z=1.3$. The formation redshift was found to be
$2.6_{-0.4}^{+0.9}$, with regard to a model cosmology where
$\Omega_m=0.3$ and $\Omega_{\Lambda}=0.7$, ignoring selection
effects. Again unfortunately, this result is based on a sample of
only 3 galaxies.

Massive cluster ETGs appear to have a large fraction
of stars that formed early in the history of the universe. It is
worth noticing that the scatter in the derived $M/L$ values
is large and confirmed by the spread in galaxy colors.

The evolution of $M/L$ ratios of ETGs has also been derived for rich
clusters at $0.02<z<0.83$
\citep[e.g.,][]{mnras281-985,apj478-l13,apj493-529}, and in the
general field out to $z\sim0.7$
\citep[e.g.,][]{mnras308-1037,mnras326-237,apj553-l39}. Cluster data
show a gradual increase of the mean $M/L$ ratio by a factor of $\sim3$
since $z=0.83$. The results are still not definitive but indicate that
field galaxies have similar $M/L$ with respect to their counterparts
in clusters. Field ETGs define a tight FP out to $z\sim0.4$, with
similar scatter as in the local sample
\citep[e.g.,][]{mnras326-237}. However, the range of the parameters
measured in field ETGs is not sufficiently extended to derive the
FP slope. The measured offset changes with redshift and seems
to be consistent with the passive evolution of a stellar population,
formed in a single burst at $z>2$.  The effective $M/L$ ratio varies
as ${\rm d}\log(M/L_B)/{\rm d}z=-0.72_{-0.16}^{+0.11}$, a little bit
faster than observed in clusters ($-0.49\pm0.05$).

The study of the FP for field ETGs can also be important to check the
predictions of the hierarchical formation model, since the stellar population
of field galaxies is expected to be significantly younger than in
clusters.

\subsection{The FP and the environment}\label{sec1c}

Several authors have tried to measure systematic differences in
the FP built with ETGs populating different clustering scales
\citep[e.g.,][]{apj389-l49,mnras257-187,apj418-72}. \citet{apj418-72}
first measured the FP for ETGs in compact groups (CGs), deriving
tentative evidences of systematic differences. They
interpreted the differences as due to lower velocity dispersion
of ETGs in CGs. The above conclusion would imply that the galactic environment
may play an important role in the process of galaxy formation and evolution.
The question has been re-examined more recently by \citet{aj122-93}, who
did not find any significant difference, as far as the FP is concerned,
between ETGs in CGs and in other environments.

The above result adds to earlier findings
\citep[][]{aj116-1591,aj116-1606} where no difference was found
between ETGs in clusters and in the field as far as the FP is
concerned.  A similar conclusion has been reached by
\citet{apj543-131} from the analysis of gravitational lens ETGs in
low-density environments.  The above mentioned galaxies lie on the
same FP of galaxies in clusters at similar redshifts.

The high-$z$ formation epoch and the lack of significant differences
between field and cluster environment, seems to contradict the
hierarchical clustering scenario. The homogeneity observed in the
properties of the FP appears to be poorly correlated with the
characteristics of the galaxy environment, implying a strong constraint
on the age of galaxies in the environments under discussion.

On the other hand, the results of the SLOAN  digital sky survey
\citep[][]{aj125-1866}, seem to indicate that the FP of galaxies
belonging to dense regions is slightly different from its counterparts belonging to less
dense regions. Of course, the large range of environments covered by the SDSS
provides a strong weight to this result. The residuals from the FP
at different redshifts show a weak correlation with the local density.
Luminosities, sizes and velocity dispersions of ETGs increase slightly
as the local density increases, while average surface brightnesses
decrease.

Along this vein, by dividing the galaxies belonging to groups into
three subsamples according to their parent X-ray luminosity,
\citet{mnras349-527} found that larger and more luminous galaxies
belong preferentially to luminous groups.

It also seems that the photometric, structural and dynamical
properties of ETGs hosting a black-hole in their center, are
substantially indistinguishable from their counterparts with no sign
of activity \citep[e.g.,][]{apj595-624,apj583-134}. BL Lac and radio
galaxies are also found on the same FP of cluster ETGs by
\citet{apj617-903}.

\subsection{The Kormendy Relation (KR)}\label{sec1d}
The KR projection of the FP in the \muere\ plane is discussed here,
since it has been often used for determining the luminosity evolution
of ETGs. In addition, it has been found useful information on physical
mechanisms involved in galaxy merging and accretions, otherwise
undetected within the framework of the classical edge-on view of the
FP correlation \citep[e.g.,][]{mnras342-501}.  Many studies have
confirmed that luminous ETGs in clusters approximately follow the
relation, $\muem = a_{KR} \log\re + const$, found by Kormendy, with a
slope, $a_{KR}\sim3$, and an intrinsic scatter, $0.3\div0.4$. However,
\citet{mnras259-323} found that once the \muere\ plane is extended to
objects fainter than $M_B\simeq-19$, it appears divided in two
families, named ``ordinary'' and ``bright''. They suggested that the
origin of such a dichotomy may be related to mechanisms of
formation and evolution.  More recently, \citet{aj125-2936} claimed
that the dichotomy between dwarf and giant ellipticals may be only an
artifact, in consequence of using (i) the de Vaucouleurs' law to fit
the galaxy profiles instead of the Sersic law, and (ii) the effective
surface brightness parameter, \muem\, instead of the central surface
brightness, $\mu_o$.

The KR for ``bright'' ETGs is considered today as a useful tool to
test their luminosity evolution. In particular, the \muere\ data for
galaxies in clusters at increasing redshift have been claimed to be
consistent with a passively evolving stellar population
\citep[e.g.,][]{Bower,Aragon,Bende1,mnras281-985,apj478-l13,JorgHj,Ziegler,apj493-529,VanDF1}. On
the other hand, some studies have also claimed that the data are
consistent with the hierarchical evolutionary scenario
\citep[e.g.,][]{Kauf2,mnras281-487}. Working with a large sample of
ETGs in three clusters at different redshifts, \citet{apj595-127}
concluded that the slope of the \muere\ relation is almost invariant
up to $z\sim0.64$ ($a_{KR}\sim2.91\pm0.08$).

Altogether, the above mentioned results seem to imply that the
mechanisms driving the evolution of galaxies in clusters do not
significantly affect the distribution of galaxies in the \muere\
parameter space. The homogeneity and the invariance with redshift of
the distributions under discussion, is also suggested by the analysis
of the SLOAN data \citep[][]{aj125-1849}.  Unfortunately, the
knowledge of the luminosity evolution of ETGs implies accurate and
homogeneous determinations of the coefficients (zero point and slope)
of the FP and the KR, both for local and intermediate/high-redshift
clusters. Concerning the latter, \citet{aa346-13} showed that the zero
point of the KR has a scatter of the same order as the luminosity
correction required from stellar synthesis evolutionary models, which
makes hard to draw any firm conclusion about the model to be
preferred.

The present ambiguity is originated by different aspects of the
FP and \linebreak\muere\ relations, which need to be clarified.
\begin{itemize}
\item The use of scaling relations to determine cosmological
parameters and luminosity evolution of ETGs, is hampered by the
problem of the metric of the Universe. \citet{apj495-l31} pointed out
that the transformation from angular radii to metric sizes
depends on the assumed world model. In consequence, luminosity
evolution and purely geometrical effects are intrinsically mixed.
\item The structural parameters of ETGs, \mue\ and \re , are generally
derived in different wavebands adopting different techniques (\eg\
measuring \re\ from the circular or elliptical aperture growth
light-curves, adopting a \rn\ or \r1q\ fit to luminosity profiles,
and using different de-convolution techniques to take into account the
effect of the seeing).
\item In principle, to get consistent and unbiased values of the
coefficients of both relationships, one should handle galaxy
samples complete either in luminosity (down to a given limiting
absolute magnitude) and cluster area coverage.
\item Nobody considers the
eventuality that an intrinsic scatter of the KR and FP coefficients
could exist even among low-redshift clusters, possibly due to different
cluster properties (richness, concentration, sub-clustering, galaxy
luminosity function, etc.) and evolution histories.
\end{itemize}

\begin{figure*}
\begin{center}
\includegraphics[bb=7cm 4cm 15cm 30cm, scale=0.6]{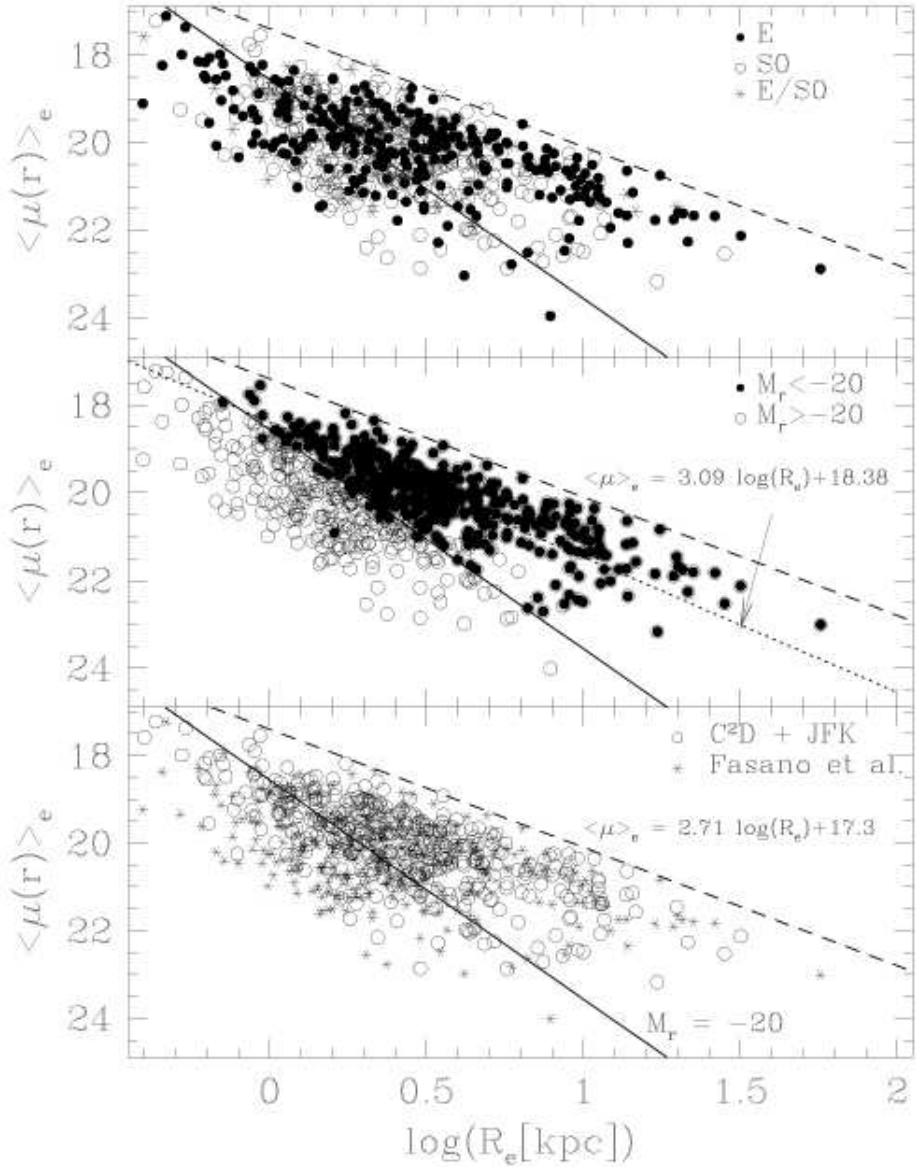}
\caption[]{Bottom panel: the distribution of ETGs from samples by
\citet{aa387-26} (asterisks) and \citet{Caon90,Caon94,mnras280-167}
(circles).  The solid line marks the locus of galaxies with equal
luminosity, $M_r = -20$ mag. The dashed line represent the LOA (see
text). Middle panel: the KR (dotted line) for the bright sample of
galaxies (filled circles). Top panel: the KR plane for different
morphological types.}
\label{figxx}
\end{center}
\end{figure*}

In order to clarify the current state-of-the-art of the KR relation,
we plot in Fig.\,\ref{figxx} the KR plane, extracted from a sample of
735 ETGs belonging to 20 nearby clusters.  The whole data set consists
of 269 ellipticals, 347 S0's, and 119 E/S0's extracted from
\citet{Caon90,Caon94,mnras280-167,aa387-26}.  The bottom panel shows
the distribution of different data samples.  The solid line is
related to fixed absolute magnitude, $M_r=-20$, in the KR plane. The
progressive absence of galaxies in the lower-left region of
Fig.\,\ref{figxx}, reflects the magnitude limits of the samples. On the
other hand, the lack of galaxies in the upper-right region of the
plot, is telling us something about the physics of galaxies. This
region of total avoidance is delimited by a line of equation, $\mue =
2.71 \log(r_e) +17.3$ (derived empirically as the linear upper
envelope, minimizing the average distance of galaxies from the line
itself). This is the ``Line of Avoidance'' (hereafter LOA), an
analogous of the ``Zone of Exclusion'' (hereafter ZOE) described by
\citet{apj399-462}, who first recognized this feature in the
$\kappa_1$-$\kappa_2$ projection of the $\kappa$-space (see section
\ref{sec1e}).

In the middle panel, we note that (i) the region occupied by the bright
ETGs in the KR plane has a conic shape, and (ii) an ideal bisector line
sharing in two parts the above mentioned region, is very close to the fit of the KR
relation (dotted line).  One can speculate that the KR may simply
arise in consequence of two selection effects: one, due to the fact
that no galaxy can live in the ``Zone of Avoidance'', and one other,
due to the selection of galaxies according to their luminosities (the
slope of the line of constant luminosity is indeed steeper than the
LOA).

It is also important to stress that the slope of the LOA is close to
that of constant central velocity dispersion, $\sigma_o$.  This leads
to suspect that the physical meaning of the LOA is essentially
related to the existence of an upper limit to the allowed velocity
dispersion of ETGs. The occurrence of a maximum $\sigma_o$ may in fact depend on the
environment wherein galaxies have been assembled, and/or on the average
temperature and density of the Universe at the epoch of formation.
Alternatively, one can hypothesize that the LOA is a fully invariant
feature of the KR plane, reflecting some universal law of
physical equilibrium.

It is obvious that investigations aimed at clarifying whether the
parameters of the LOA depend or not (and, if it is the case, how do they
depend) on both the environment (field/cluster, global cluster
properties) and the redshift, would be of great interest in order to
understand the processes involved in the formation and evolution of
ETGs. It will be soon possible thanks to the efforts of
WINGS (Wide-field Imaging of Nearby Galaxy cluster Sample) survey by \citet{Fasano04}.

\subsection{The Photometric Plane}\label{secxx}
During the last decade, it has been proved that the light distribution
of ETGs very closely follows the generalized de Vaucouleurs' law
proposed by \citet{sersiclaw}. Accordingly, the radial luminosity profile
of a galaxy reads:
\begin{lefteqnarray}
I(r)=I(0)\exp[-b_n(r/r_e)^{1/n}]~~;
\end{lefteqnarray}
where $I(0)$ is the central intensity, \re\ the effective radius, and
$b_n$ is a function of the shape parameter, $n$, chosen to scale the
radius, \re , to enclose half of the total luminosity. An equivalent
expression is $L_T=k_L\iem r_e^2$, where the parameter $k_L= 2\pi n
(e^{b_n}/b_n^{2n}) \Gamma(2n)$, depends on the shape of galaxy
light distribution.  The shape parameter, $n$, is related to both the
curvature of light profiles and to the degree of concentration of
luminous matter in ETGs and spiral bulges. The concentration index is, in
fact, an additional important parameter to be used in characterizing
galaxian structures.

With regard to ETGs, \citet{mnras265-1013} found a correlation between the Sersic
index, $n$, and the logarithm of the effective radius, \re , and
between $n$ and the total galaxy luminosity.
A correlation between $n$ and $\sigma_o$ was also been found by
\citet{aj121-820}.   \citet{apj531-l103} showed that a tight
correlation exists in the Sersic parameter space
defined by the set of coordinates, $[\log\re, \mu_o, \log\nu (=1/n)]$,
with regard to 42 Coma ellipticals and 26 spiral bulges
in the Near-IR band.  The result is:
\begin{equation}
(0.173\pm 0.025) \log\re - (0.069\pm 0.007) \mu_o=-\log\nu-(1.18\pm 0.05)~~;
\end{equation}
which makes the so called {\it Photometric Fundamental Plane} (PFP).
Working with the B-band data set of \citet{Caon94},
\citet{mnras334-859} obtained a similar relation:
\begin{equation}
\log\re \propto (0.86\pm0.13) \log n - (1.42\pm0.22) \muem~~;
\end{equation}
which shows a scatter in $\log(\re)$ of $38\div48\%$ per galaxy.
Though the PFP has the advantage of avoiding velocity dispersion
measurements, still some uncertainties remain with regard to
high redshift galaxies, due to difficulties in determining
the value of $n$ when spatial resolution decreases.

A theoretical interpretation of the PFP and related scaling laws has
been tempted by \citet{mnras309-481} and \citet{aa353-873,aa379-767}.
After a violent relaxation phase, an elliptical galaxy is assumed to
reach a stage in which the entropy of the star fluid is
quasi-stationary. The location of E galaxies in the Sersic space can
be related to the intersection of two particular surfaces: one, of
constant specific entropy (which weakly depends on mass), and one
other, of constant galaxy potential energy. The related intersection
is defined as the {\it Entropy-Energy line}.  The consequence is that
all galaxies have roughly the same specific entropy, obey the same
energy-mass relation, and the Sersic law. From the photometric point
of view, E galaxies appear as a one-parameter family, described \eg\
by $\nu$, which is small for giant objects and large for dwarf
objects. Some further comments on this approach will be considered in
section \ref{teoria}.

\subsection{The $\kappa$-space}\label{sec1e}

\citet{apj399-462} introduced a new orthogonal coordinate system
known as the $\kappa$-space, defined by the following independent
variables:
\begin{lefteqnarray}
&&\kappa_1=(\log~\sigma_0^2+\log~r_e)/\sqrt{2}\sim\log~M \nonumber ~~;\\
&&\kappa_2=(\log~\sigma_0^2+ 2\log~\iem-\log~r_e)/\sqrt{6} \sim \log~ \left( \frac{M}{L}\iem^3 \right)~~; \\
&&\kappa_3=(\log~\sigma_0^2- \log~\iem-\log~r_e)/\sqrt{3} \sim \log~\frac
{M}{L}~~; \nonumber
\end{lefteqnarray}
which relate to galaxian total mass, $M$, average effective surface
brightness, $\iem$, and mass to light ratio, $M/L$, respectively.

The above mentioned parametrization of the FP is useful
for a number of reasons, namely (i) $\kappa$-variables are
expressed only in terms of observables, independent of specific
models or assumptions; (ii) the $\kappa_1$-$\kappa_3$ plane represents an
edge-on view of the FP and provides a direct view of the tilt; and (iii)
the $\kappa_1$-$\kappa_2$ plane almost represents a face-on view of the
FP.

\begin{figure*}
\begin{center}
\includegraphics[bb=7cm 10cm 15cm 30cm,scale=0.6]{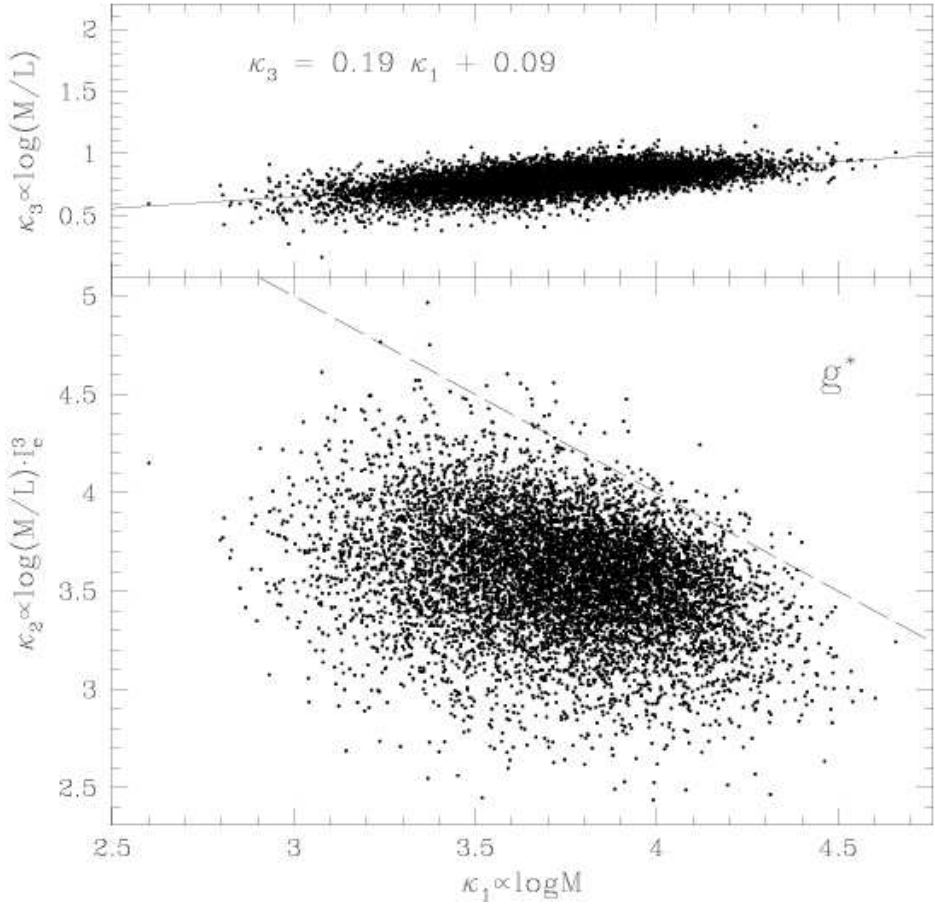}
\caption[]{The $\kappa$-space representation for $\sim 9000$ galaxies
  from the SLOAN survey in the $g^*$ band \citep[for further details
  see][]{aj125-1866}. The dashed line is the Zone of Exclusion (ZOE), expressed as
  $\kappa_1+\kappa_2=8$.}
\label{kappaspace}
\end{center}
\end{figure*}

The SLOAN $g^*$ band data in the $\kappa$-space are shown in
  Fig.\,\ref{kappaspace}.  The most important features are the {\em
  tilt}, the {\it tightness}, i.e. the nearly constant and very small
  dispersion of $\kappa_3$ at every location on the FP, and the ZOE,
  which acts as an upper limit on the location of galaxies on the
  $\kappa_1$-$\kappa_2$ plane.

The $\kappa$-space parametrization has been strongly criticized by
\citet{aj116-1591}. They first argued that the variable, $\kappa_1$,
attains different values for different wavelengths via the dependence
on the effective radius, $r_e$, which implies different masses for
different wavelengths, contrary to what is expected.  A second problem
comes from the observation, that lowest luminosity ellipticals do not
show detectable color gradients \citep[][]{aj100-1091} suggesting
there could be a dependence of color gradients on luminosity (and
hence on mass). Accordingly, both $\kappa_1$ and $\kappa_3$ show a
dependence on wavelength.

Despite the above mentioned criticism must be taken into consideration,
we recognize that the $\kappa$-space is very useful for testing
predictions of galaxy formation and evolutionary models.  In fact,
assuming that only small deviations from homology are possible (\eg\ in
ETGs), the $\kappa$-space has the advantage that the processes involved
in galaxy formation and evolution, can directly be tested by observing
the shift of galaxy points in the related frame.  For example,
\citet{mnras335-335} tested their $N$-body-tree-SPH model simulations of
elliptical galaxies by comparison between model and observed
galaxy distribution in the $\kappa$-space.  They found
that model galaxies with different SF history, from small to large
ellipticals (producing different mean ages for the bulk of stellar
population), are in quite good agreement with observations.

\section{The tilt of the FP}\label{sec2}
Let us summarize different investigations aimed to explain the FP
tilt.  In the B-band $\kappa$-space representation,
\citet{apj114-1365} have given the following relation for the tilt
(updated results are shown in Fig.\,\ref{kappaspace}, top panel):
\begin{lefteqnarray}
\label{kappatilt}
&&\kappa_3=0.15~\kappa_1 +0.36\Longrightarrow M/L\propto
M^{\alpha_t}~;~\alpha_t\simeq 0.2~~.
\end{lefteqnarray}

The reasons of the deviation from model predictions discussed
above, are not yet well established, although several
explanations have been proposed in the recent past.

\subsection{The role of anisotropy}\label{sec2a}
The effects of radial orbital anisotropy in the velocity distribution
of ETGs has been discussed in detail by
\citet{mnras332-901}. According to their results, anisotropy
contribution is only marginal.  Given an isotropic galaxy model lying
on the FP, its counterparts where radial orbital anisotropy is
sufficiently high, are found to move away.  But, at the same time,
instabilities set on, and the related end-products fall back again on
the FP.

In the conclusions of the authors, a systematic increase of radial
orbital anisotropy with galaxy luminosity cannot explain, by itself,
the whole tilt of the FP.  The above result has been obtained using
$N$-Body simulations together with spherically symmetric, both
one-component $\gamma$ models, and two-component ($\gamma_1,\gamma_2$)
models.  Accordingly, the density profiles obey the law:
\begin{lefteqnarray}
\frac\rho{\rho_0}=f(\xi)=\frac{C_{\rm norm}}{\xi^\gamma
(1+\xi)^{4-\gamma}}~~,~~ \xi=\frac{r}{r_0}~~;~~0\leq \gamma \ <3~~;
\label{gammamodel}
\end{lefteqnarray}
where $C_{\rm norm}$ is a normalization constant. The models under
discussion depend on five parameters, namely a scaling radius, $r_0$,
a scaling density, $\rho_0$ (or the total mass, $M$), the stability
parameter, $s_a=r_a/r_0$, \ie\ the ratio of anisotropy
radius\footnote{The anisotropy radius, $r_a$, is very important
as the velocity dispersion tensor is mainly radially anisotropic
at sufficiently large distances ($r\gg r_a$), nearly isotropic at
sufficiently small distances ($r\ll r_a$), and perfectly isotropic at
the center ($r\to0$).  Accordingly, the velocity dispersion tensor
tends to be globally isotropic in the limit, $r_a\to+\infty$ . For
further details, see \citet{sval5-42,aj90-1027,apj520-574,mnras332-901}.}  to scaling
radius for B component, the mass ratio of D to B component, $m=
M_D/M_B$, the ratio of D to B component scaling radius,
$\beta'=(r_0)_D/(r_0)_B$.

Taking into consideration the tilt expressed by Eq.(\ref{kappatilt}),
under the assumption that present-day ETGs are homologous and
virialized systems, the following relation holds:
\begin{lefteqnarray}
\frac{G L \Upsilon_*}{\re} = K_V \sigma^2_0~~;
\label{eq:dinamic}
\end{lefteqnarray}
where $\Upsilon_* = M_*/L$ is the stellar $M/L$ ratio and $K_V$ is a
dimensionless factor depending on the stellar density profile, the
internal kinematics, the DM amount, and the distribution and
relative orientation of the galaxy with respect to the observer's line
of sight \citep[see][]{aa321-724}.  In terms of mass and luminosity,
Eq.\,(\ref{eq:dinamic}) may be cast under the form \citep[][]{aa321-724}:
\begin{lefteqnarray}
\frac{\Upsilon_*}{K_V} \simeq 1.12 \cdot L^{0.23}~~. \label{pianofon}
\end{lefteqnarray}

It is apparent that
the trend of $M/L$ with luminosity depends on the properties of
the parameter, $K_V$.   Given a set of $\gamma$-models which
differ only in the value of the stability parameter, $s_a$, a
threshold, $s_{as}$, is found to exist, below which $(s_a<s_{as})$
initial configurations evolve towards radial instability.
Accordingly, the initial configuration related to the threshold
value, virializes at a maximum distance from the FP via the
$K_V$-$s_a$ dependence.   Initial configurations below the
threshold can attain larger distances from the FP, but move
back to the FP after virialization following radial instability.

With regard to one-component models, \citet{mnras332-901} obtained
that the end-products, with radial unstable initial conditions, are
strongly asymmetric, and their representative points in the
$\kappa$-space span a range of values which depend on the line of
sight orientation. Further analysis shows that, owing to the above
mentioned effect, not only the FP tightness is nicely related to
stability, but the FP -by itself - acts as an ``attractor'' for the
end-products of unstable systems, provided initial configurations are
placed therein.

With regard to two-component models, a less flattened final
stellar distribution has been found especially with massive
DM halos. In particular, there is evidence that two-component models
with light halos have a similar behavior with respect to
one-component models, but
massive DM halos tend to prevent the end-products of unstable
models to fall back on the FP, the effect being larger for more
concentrated halos.

One can now examine whether Eq.(\ref{pianofon}) can be satisfied with
a systematic variation of $K_V$ resulting from an underlying
correlation of the stability parameter as a function of
galaxy luminosity, $s_{a}=s_{a}(L)$, model structure,
and $\Upsilon_*$.

The FP tilt over the whole observed range ($\Delta \kappa_1
\simeq 3$ and $\Delta \kappa_3 \simeq 0.4$, see Fig.\,\ref{kappaspace})
cannot be reproduced in dealing with one-component models.  In fact,
independently of the value of $\gamma$, the allowed variations in $K_V$
correspond to $\Delta \kappa_3 \leq 0.04$, much smaller than the
observed range.
If unstable (but consistent) models are also taken into account, the
$K_V$ range increases up to allow $\Delta \kappa_3 \leq
0.2$, which is yet significantly smaller with respect to its
counterpart deduced from observations.

With regard to two-component models, DM halo concentration plays an
important role in the displacement of end-products.
For each family of initial conditions, under the sole
requirement of consistency, the end-products yield $\Delta \kappa_3
\leq 0.2$, again smaller than the observed range \citep[][]{mnras332-901}.

\subsection{The role of weak-homology}\label{sec2b}
If the $n$-$L$ correlation discussed in section \ref{secxx} is real and
physical, the luminosity profile of ETGs varies according to
their size: larger galaxies are more centrally concentrated than
smaller ones. In other words, ETGs are not homologous stellar systems,
and one of the key assumptions in the interpretation of the FP, \ie\
the structural homology, no longer appears to be true.

\citet{aa386-149} studied a small set of objects characterized by
photometric profiles that deviate significantly
from the standard \r1q\ law.
They argued that ETGs might be considered as weakly homologous systems, on
the basis of the observed deviations from the \r1q\ law and the
existence of a $n$-$L$ correlation.

The normalized mass-luminosity ratio, $\Upsilon_*/K_V$, no matter how
complex the galaxy structure is, turns out to be a well defined
function of any two of the observables ($L$, \re, $\sigma_o$).  The
substitution of the empirical $\kappa_1$-$\kappa_3$ relation into
Eq.\,(\ref{eq:dinamic}) allows the following expression:
\begin{lefteqnarray}
\frac{\Upsilon_*}{K_V} \propto \re^{(2-10B'+A)/A}
L^{(5B'-A)]/A}~~,
\label{upsi}
\end{lefteqnarray}
in terms of $r_e$ and $L$, where $A$ and $-2.5B'=B$ are the FP
coefficients appearing in Eq.(\ref{eq:pianofond}).   Equivalent
expressions in terms of $r_e$ and $\sigma_o$, or $L$ and $\sigma
_o$, could also be derived.

Accordingly, while the whole set of stellar systems described by
Eq.(\ref{eq:dinamic}) are in virial equilibrium, only the subset for
which $\Upsilon_*/K_V$ scales according to Eq.(\ref{upsi}) are placed
on the FP.  Since the exponent of \re\ is very small ($\sim0.04$ in
the B-band), the restriction, $2-10B'+A=0$, allows to explore two
different extreme cases, namely (i) strict homology (\ie\ $K_V$
identical for all galaxies), which implies $\Upsilon_* \propto
L^{\delta}$, where $\delta= (2-A)/2A \simeq 0.30 \pm 0.064$; and (ii)
constant $\Upsilon_*$ and weak homology, which implies $K_V$ is a well
defined function of \re\ and $L$ ($K_V \propto L^{-\delta}$).

The latter alternative seems to be qualitatively encouraged by
observations; in fact the observed excursion, $1\leq n \leq 10$, is
consistent with the desired range of $K_V$. However, a fine-tuning
problem remains for changes in $K_V$.

In conclusion the ranges of variations in $\Upsilon_*$ and $n$ at
fixed galaxy luminosity, can be substantial and consistent with an
arbitrarily thin FP, but the involved fine-tuning requires a
$\Upsilon_*$-$n$ conspiracy, in the sense that changes in $\Upsilon_*$
are compensated by special changes in $n$ and vice versa.
In the conclusions of \citet{aa386-149}, neither strict homology
nor weak homology scenario explain the observed FP in
a satisfactory and conclusive way.

According to recent investigations \citep[][]{apj600-l39}, the FP tilt
is due for about three fourths to structural non homology and for
about one fourth to stellar population effects. In the conclusions of
the authors, even very massive ETGs are strongly dominated by DM in
their inner regions.

Questioning whether the central velocity dispersion accurately
represents the kinetic energy of a galaxy, \citet{symp97}
suggested that the
contribution of rotation to total kinetic energy in ETGs is not
negligible, and rotation is not correlated to velocity
dispersion. In the finding of the authors, the kinetic energy of
random motions is
proportional to $\sigma_o^{1.6}$ instead of $\sigma_o^{2}$, and
the shapes of velocity dispersion profiles show a definite trend,
where the larger gradient is related to the larger
$\sigma_o$.

The above mentioned result
seems to favor a dissipative scenario for the formation of ETGs, in
which ``hotter'' systems underwent more effective {\em cooling}.
In this view, most of
the FP tilt ($\sim 55\%$) seems to arise from a sort of
dynamical non-homology of the systems, \ie\ systematic
changes of the relations between observed ``local'' quantities
used to define the FP, and physical ``global'' quantities
appearing in the formulation of the virial theorem.

\subsection{The role of IMF and SF}\label{sec2c}
As shown above, if ETGs are homologous and isotropic systems,
the FP tilt may simply be attributed to variations of $M/L$
ratio with mass, due to differences in stellar content and IMF.
Since current semi-analytical models of ETGs, based on
supernova-driven galactic winds and the usual recipes for the IMF, are
still unable to explain the FP tilt, \citet{aa339-355} explored the
idea of an IMF varying with the physical conditions of the interstellar
medium. The related model seems able to explain the tilt, leaving aside
additional effects.

An alternative viable explanation, investigated by
\citet{mnras335-335}, is a SF which depends on the mass of the system,
where SF starts later (or has a longer duration) in low-mass galaxies,
allowing significant variations in mean stellar age with respect to
high-mass galaxies.

By exploring the possibility of a systematic change of IMF along
the FP, an acceptable explanation of the FP tilt has been found, provided a
substantial variation of the IMF is allowed \citep[][]{apj416-l49}.
In addition, the constant thickness of the plane induces unreliable
small dispersions in IMF.

Let us define a power-law IMF:
\begin{lefteqnarray}
&& \phi (M_i) \propto L_B M^{-(1+x)}_i~~;
\end{lefteqnarray}
where $L_B$ is the total
luminosity in B-band.  Accordingly, the $M/L$ ratio is:
\begin{lefteqnarray}
&& \frac{M}{L_B} = 2.9 \int_{M_{inf}}^{100} M_i  M^{-(1+x)}_i dM_i~~;
\end{lefteqnarray}
where $M_{inf}$ and 100 are the lower and upper stellar mass limit in
solar units, respectively.  Two cases were considered in the above
mentioned paper, namely constant and variable power-law IMF exponent,
$x$. In the former alternative, the tilt can be reproduced at the
price of an unreasonably high lower stellar mass limit, $M_{inf}$. In
the latter alternative, a variation of at least $0 \leq x \leq 1.7$ is
needed. Since $M/L$ has a non linear dependence on $x$, one would
expect a larger dispersion in $M/L$ for increasing $x$, something like
a cone shape distribution along the FP, which is not observed.

By considering synthetic stellar $M/L$ ratios for
single burst stellar populations (in different photometric bands as
functions of age and chemical composition), \citet{asp163-28}
explored the dependence of
$M/L$ on IMF slope.  An important effect on metallicity
results: $M/L$ increases as metallicity increases. This pattern
is evident in U band but disappears in the near-IR.  In author's
conclusion, if less bright ETGs exhibit [Fe/H]$ \sim -0.3$, then metallicity
explains approximately $68\%$ of the observed tilt. It is clear,
however, that the above mentioned
effect has no influence on near-IR passbands, and a different
interpretation is needed to explain the tilt to this respect.

\subsection{The role of Dark Matter}\label{sec2d}
The role of DM on the FP tilt has been investigated by
\citet{mnras282-1}. The investigation tried two different approaches, one
ascribing the tilt to a trend in the dark-to-bright mass ratio,
$m=M_D/M_B$, at constant $\beta'=(r_0)_D/(r_0)_B$, and the other vice
versa.  Considering different combinations of empirical profiles
representing D and B components, both approaches succeed in explaining
the tilt, but a fine tuning is again necessary to reproduce a very
small scatter in the FP.  The latter alternative is preferred, being
ETGs characterized by high surface brightness and stellar density in
presence of low effective radii, at the faint end of the FP.  Then the
above mentioned galaxies show a larger concentration in B
component with respect to D component, while the effect is reduced in
their counterparts lying on the other end of the FP.

Studying the origin of the small scatter of ETGs around the FP,
\citet{mnras341-1109} claimed that inside \re , luminous matter
must largely dominate over DM halo in order to keep objects close
enough to the FP. In particular, cuspy DM distributions, as predicted
by numerical simulations in $\Lambda$CDM cosmologies, seem to be in
contradiction with the existence of the FP (see section \ref{cuspycore}). The structural properties
of dark and luminous matter are interwoven i such a way to produce a
curved surface in the space ($\log\sigma_o$, $\log\,$\re, $\log L$),
instead of a plane. In order to maintain a small scatter in the FP, it
is needed either unacceptably low, total dark-to-luminous mass ratios,
or DM halo concentrations within the range, $5\div9$, below values
related to current predictions. In addition to spiral and dwarf
galaxies, cored DM halos where density decreases as $r^{-3}$ at
larger radii, are found to be surprisingly successful in explaining
ETGs structure, pointing to an intriguing homogeneous scenario of
galaxy formation for all morphological types.

\subsection{The role of Clausius' virial maximum}\label{teoria}
In a series of papers
\citep[][]{na5-403,na6-339,na8-629,secval,na2005} the role of the
Clausius' potential energy is taken into account within the framework
of a dynamical explanation to the FP.  The analysis has been carried
out by the powerful tool of tensor virial theorem
\citep[][]{apj130-414,apj158-L141,gd87-BT}, extended to two-component systems
\citep[e.g.,][]{efe69-C,brosche83,aa139-411,apj395-119,apj528-l5,an325-326}.
Two-component models characterized by different power-law density
profiles involving homogeneous cores, have been used for similar and
similarly placed spheroids.  The outputs of this kind of models were
summarized and compared with some observable scaling relations for
pressure-supported ellipticals and, in general, for two-component
virialized systems.

In this framework, a special (virialized) configuration is identified,
and its occurrence is interpreted as a physical reason for the
existence of a FP for ETGs.  Clausius' virial energy is maximized by
the configuration under discussion, and the related radius ({\it tidal
radius}) is claimed by the authors to work as a scale length induced
by the tidal action of D on B. The above mentioned choice makes a
further constraint among physical parameters, and allows to reproduce
the exponents, $A$ and $B$, of the FP. In addition, the {\it tilt}
($\alpha_t\simeq 0.2$) comes from the \iem-\re\, relation linked to a
fixed cosmological scenario, instead of different amount of DM from
galaxy to galaxy or breaking the {\it strict homology}.  For assigned
values of B and D mass, the special configuration allows
two-dimensionality scale relations for each of the three quantities:
$\sigma_o$, \iem, \re.

It is worth of note that the occurrence of a maximum in the Clausius' virial
is not an universal property of two-component systems, but depends on mass ratio and
density profile.
Using the model discussed above, the best fit to the FP takes place
for a D to B subsystem total mass ratio of about ten.
If much lower values were established or deduced from observations, the related model
would provide no viable interpretation to the FP.

From the analysis of the dynamic and thermodynamic properties, the
authors deduce that the above defined {\it tidal radius} works as a
confinement for the stellar subsystem, similarly to the {\it tidal
radius} induced on globular clusters by the hosting galaxy, as
\citet{hoerner} found for the Milky Way. The new result appears as a
general extension of the old one to the case of concentric structures.
It could add further insight to the fact, that different kinds of
astrophysical objects, with a completely different formation history,
but subjected to a tidal potential - and then characterized by a {\it
tidal radius} - lie on the same FP
\citep[][]{apj438-L29,apj114-1365,secco2003}.  It should be noted that
a tidal radius induced by a tidal potential appears to the authors
very similar to the truncation, which \citet{aj71-64} introduced {\it
ad hoc} in his primordial models for ellipticals, extrapolating known
data for globular clusters.  In addition, the exponents, $A$ and $B$,
and the parameter, $\alpha_t$, related to the FP, are found to depend
only on the inner, universal DM distribution, implying that other
families of galaxies or, in general, astrophysical virialized objects,
necessarily belong to a similar FP \citep [as observed in the {\em
cosmic meta-plane} defined by][]{apj114-1365}.

The problem of extending the results from power-law to \citet{mnras278-488}
density profiles, which provide a closer fit to elliptical galaxies,
is also taken into consideration.

The dynamical approach outlined above diversifies from some other
apparently similar approaches
\citep[e.g.,][]{mnras309-481,aa353-873,aa379-767}.  In the latter, the
stationarity of stellar ``gas'' entropy is assumed in a pragmatic way
without trying a derivation from first principles. With regard to the
former, the DM presence is fundamental on two respects.  First, it
provides a real scale length to the gravitational
field induced by the stellar subsystem.  Second, it makes possible to
deduce, on the basis of mechanical and thermodynamical main
principles, how the above mentioned scale length maximizes the entropy
of the stellar subsystem.

\subsection{The {\it cusp/core} problem}\label{cuspycore}

Last but not least, the problem we wish to address concerns the
connection between a dynamical explanation to the FP and DM
distribution within the inner part of halos (e.g., the effective
radius \re), where most of baryons are located.  Current,
high-resolution, collisionless numerical simulations both in standard
CDM and concordance $\Lambda$CDM cosmological models, yield a general
consensus about a {\it cuspy} density halo profile (at the smallest
radii probed by simulations) i.e. $\rho(r)\sim r^{-\alpha_{int}}$,
where $\alpha_{int}$ is close to unity. To this respect a number of
discrepancies have been found $\alpha_{int} <1$
\citep[e.g.,][]{sub00,tn01,ricc03}, and $\alpha_{int}=1.5$
\citep[e.g.,][]{fm97,fm01,moo98,moo99}. On the other hand,
\citet{hay04,nav04,pow03}, concluded, by using a suite of high
resolution simulations, that $\alpha_{int}=1$ remains consistent with
simulated halo shapes (they found mean values $\alpha_{int}=$
[$1.1$;$1.2$;$1.35$] respectively), while they claim that cusps as
steep as $\alpha_{int}=1.5$ are ruled out \citep[][]{spek05}.

On the other hand, many observations, by gravitational lensing for
 galaxy cluster scale \citep[e.g.,][]{sand04}, and long-slit
 spectra for dwarf galaxy scale \citep[e.g.,][]{spek05}, suggest
 that real halos exhibit {\it cores} instead of {\it cusps}. In
 particular, \citet{salu01} concludes that halos have cores larger
 than the corresponding disk scale lengths, from a robust analysis of
 137 disk-dominated galaxies.

The related, apparent contradiction makes the so called {\it cusp/core}
problem. A brief, good review on these topics may be found in the
introduction of \citet{spek05}.  The results of the theoretical
approach presented in the previous section do not agree with the
conclusions of \citet{hay04,nav04,pow03}, but are consistent with the
observations, which are seriously challenging the {\em cuspy} profile.

A recent investigation \citep[][]{nav04} exploits the possibility of a
density profile where the logarithmic slope decreases inward gradually
until a minimum value is attained at the center. Accordingly, DM halos
can be interpreted as Sersic models \citep[][]{merrit2005}.

Another enhancement to this respect comes from the papers of
\citet{phdgar-03} and \citet{garr04}.  Using high-resolution rotation
curves of spiral galaxies obtained by Fabry-Perot spectroscopy, and
modeling DM halos using {\it Zhao} density profiles, the authors have
shown a very strict limit for the inner slope, $\gamma\le
0.8$. Further analysis on the fit parameters of \citet{phdgar-03}
sample from a physical point of view, shows that two-component models
where DM halos with different $\gamma$ are selected, remain consistent
with both CDM and $\Lambda$CDM cosmologies \citep[][]{bin05}.
Preliminary conclusions for the sample considered above, agree with
\citet{salu01} results ($\alpha_{int}<1$, till the disk scale length).
It should be noted that the order of resolution ranges from hundredths
to thousandths of DM virial radius, both for observations and the
corresponding {\it fit} models, which results to be comparable to, or
better than, the best resolved simulations.  On the basis of pure
theoretical grounds (the Jeans' equation together with reasonable
assumptions on the gravitational potential), \cite{aa404-809}, claim
that asymptotic inner slopes of DM halos cannot lie outside the range,
$0\le\gamma\le 0.5$.  The debate is still open, and further research
is needed on this subject.

\section{Conclusions}\label{sec4}
We have reviewed some of the more relevant observational
investigations on the FP, and summarized the most important
theoretical attempts to explain its tilt and tightness by
means of velocity anisotropy, DM fraction, homology, and IMF. None of
the above mentioned mechanisms, by itself, is currently able to
provide a complete explanation of the observed properties of the FP.

We have shown that starting from the tensor virial theorem extended to
a two-component model for ETGs, more insight on the problem can be
gained.  If a maximum of Clausius' virial energy is really the key for
a dynamic explanation of FP, it appears surprising how the
corresponding special configuration links together many of the
involved parameters, in such a way that the basic two-dimensionality
of the manifold of ETGs is ensured.  The main observables \re\,, \iem\
, and $\sigma_o$, appear to depend on cosmology via $\gamma'$ (a
quantity related to the local slope of the {\it mass variance})
implying the projections of the FP (\eg\, FJ, KR) also depend on it.
Nevertheless, the FP equation is found to degenerate with respect to
$\gamma'$.

The debate on the origin of the FP is still currently open, but
interesting developments are to be expected as soon as larger field
surveys will be available in the near future. \\

{\bf Acknowledgments}
This work has partially been supported by {\em Fondazione Cassa di Risparmio
  di Padova e Rovigo}, Piazza Duomo 15, Padova (Italy). We thank an anonymous referee for critical remarks.

\end{document}